\documentclass[aps,pre,twocolumn,showpacs,groupedaddress]{revtex4}

\usepackage{graphicx,psfrag}

\def\BibTeX{{\rm B\kern-.05em{\sc i\kern-.025em b}\kern-.08em
    T\kern-.1667em\lower.7ex\hbox{E}\kern-.125emX}}

\begin{document}

\title{Uniform framework for the recurrence-network analysis of chaotic time series}

\author{Rinku Jacob}
\email{rinku.jacob.vallanat@gmail.com}
\affiliation{Department of Physics, The Cochin College, Cochin-682 002, India}
\author{K. P. Harikrishnan}
\email{kp_hk2002@yahoo.co.in}
\affiliation{Department of Physics, The Cochin College, Cochin-682 002, India} 
\author{R. Misra}
\email{rmisra@iucaa.ernet.in}
\affiliation{Inter University Centre for Astronomy and Astrophysics, Pune-411 007, India} 
\author{G. Ambika}
\email{g.ambika@iiserpune.ac.in}
\affiliation{Indian Institute of Science Education and Research, Pune-411 008, India} 

\begin{abstract}
We propose a general method for the construction and analysis of 
unweighted $\epsilon$ - recurrence networks from chaotic time series. The selection of the 
critical threshold $\epsilon_c$ in our scheme is done empirically and we 
show that its value is closely linked to the embedding dimension $M$. 
In fact, we are able to identify a small critical   
range $\Delta \epsilon$ numerically that is approximately the same for the random 
and several standard chaotic time series for a  
fixed $M$. This provides us a uniform framework for the non subjective comparison of 
the statistical measures of the recurrence networks constructed from various 
chaotic attractors. We explicitly show that the degree distribution of the 
recurrence network constructed by our scheme is characteristic to the structure of the 
attractor and display statistical scale invariance with respect to increase in the 
number of nodes $N$. We also present two practical applications of the scheme, 
detection of transition between two dynamical regimes in a time delayed system and 
identification of the dimensionality of the underlying system from real world data 
with limited number of points, through recurrence network measures.  
The merits, limitations and the potential applications of the proposed 
method have also been highlighted.
\end{abstract}


\maketitle

\section{\label{sec:level1}INTRODUCTION}
Study of networks has become an important area 
of research in the last two decades cutting across various disciplines and often providing a coherent view of 
structures and phenomena which may appear different from a common perspective. Mathematically, networks are 
entities defined on an abstract space, with $N$ number of nodes and arbitrary number of links  
between them. We often refer to \emph{complex networks}, implying that the structure is irregular and complex.  
Such networks may be weighted or unweighted, directed or undirected depending 
on the structure and interaction of the system it tries to model. 

Historically, the study of networks has been the domain of a branch of mathematics called 
\emph{graph theory}. The mathematical basis for the analysis of complex networks was laid using the so called 
``random graphs''(RG) by Erd\H os and R\' enyi about five decades back \cite {erd}. They have shown several 
properties for random graphs, which are now popularly known as Erd\H os-R\' enyi (E-R) networks. 
The measures introduced for the analysis of RGs have, naturally, become the tools to characterize the 
structure and evolution of complex networks. The most important among them are the degree distribution and 
the characteristic path length (CPL), apart from other measures, such as, link density (LD), clustering 
coefficient (CC), diameter, centrality, etc. The degree distribution indicates how many nodes $n_k$ among the 
total number of nodes $N$ have a given degree $k$. It is usually represented as a probability distribution 
$P(k)$ as a function of $k$, where $P(k) = {{n_k} \over {N}}$. For E-R networks, one can show that $P(k)$ 
follows a Binomial distribution which, for large $N$ and small probability of connection tends to the 
Poisson distribution. The CPL, denoted by $<l>$, is defined through the shortest path $l_s$ connecting 
two nodes $\imath$ and $\jmath$. For unweighted and undirected networks that we consider in this work, 
$l_s$ is defined as the minimum number of nodes to be traversed to reach from $\imath$ to $\jmath$. 
The average value of $l_s$ for all the pair of nodes in the whole network is defined as $<l>$ and the 
maximum value of $l_s$ is taken as the diameter of the network, denoted by $l_D$. 
For a detailed discussion of all the network measures, 
see the popular books by Newman \cite {new1} and Watts \cite {wat1} and some excellent reviews 
on the subject \cite {alb1,boc}.

An important area where the new network based concepts and measures have been applied successfully is in the 
analysis of dynamical systems \cite {str1,barz}, especially those showing chaotic behavior, by constructing 
complex networks from time series data of the dynamical systems.  
Here, the aim is to extract information regarding 
the structure of the underlying chaotic attractor, which are otherwise difficult to get using the 
conventional methods of nonlinear time series analysis. 
The basic idea of this technique is that the information inherent in a chaotic time series is mapped on to 
the domain of a complex network using a suitable scheme. One then uses the statistical measures of the complex network 
to characterize the underlying chaotic attractor. 

Two questions are relevant in this context. Firstly, 
which method is to be used to transform the time series into the corresponding network and secondly, how to  
ensure that the resulting network truly represents the characteristic features of the underlying attractor.  
To answer the first, several methods have been suggested in 
the literature, such as cycle networks \cite {zha1}, visibility graphs \cite {lac}, transition networks 
\cite {nic} and recurrence networks (RN) \cite {mar1}. The RNs can be constructed in different ways, 
namely, the correlation networks \cite {yan1}, $k$-nearest neighbour networks \cite {xu1} and 
$\epsilon$-recurrence networks \cite {don1}. It has been shown that the networks generated by each of 
these methods can capture several characteristics of the chaotic time series like dynamical 
transitions in the system, topological properties of the attractor, etc. Hence each method is relevant 
in the context of specific applications. A detailed discussion of these methods and their comparison can be 
found elsewhere \cite {don2,don3}. 

Among the methods mentioned above, the one based on $\epsilon$ - recurrence is 
physically appealing since it is based on the concept of \emph {recurrence} of a trajectory in 
the phase space. Moreover, the method can be applied to any type of synthetic or real world data and 
the resulting networks are found to be useful tools for uncovering complex bifurcation scenario and detecting 
dynamical transitions in palaeo-climate data \cite {zou1,dong1,gao1}. The methods based on recurrence have also 
found diverse applications 
ranging from life sciences \cite {wes1,ram1}, earth sciences \cite {tra1} and astrophysics \cite {zol}. 
In this work, we concentrate on 
this method for network generation, confining ourselves to the case of unweighted 
$\epsilon$-recurrence networks. 

The answer to the second question raised above leads us to the choice of the 
parameters for network construction.
A crucial parameter in the construction of the RN is the recurrence 
threshold $\epsilon$ itself. In all the existing schemes, the value of $\epsilon$ chosen is different for 
each time series, whatever be the criteria used for its selection, due to the arbitrary size of the 
attractor after embedding. Our main goal in the present work is to propose a scheme that uses an 
approximately identical range of values for $\epsilon$  for different time series, both synthetic 
and real world, for a given embedding dimension.  Apart from providing a uniform framework for  
the recurrence network analysis, the scheme has several advantages 
and practical applications as discussed below in detail.  

Our paper is organised as follows: In the next section, we discuss the basic idea of RN construction  
while in \S III,  the criteria for the selection of 
all the parameters for the construction of RN from the time series are presented.   
We then proceed, in \S IV, to construct the RNs from several low 
dimensional chaotic systems. All the important network measures are derived from the RNs as a 
function of $M$ and $N$ and compared. The degree distribution, especially, is studied in detail 
and is shown to be characteristic of the structural complexity of a chaotic attractor. Two practical 
applications of the proposed scheme are  illustrated in \S V. 
A discussion on various aspects of implementation of the scheme and conclusions are given in \S VI. 

\section{\label{sec:level1}CONSTRUCTION OF RECURRENCE NETWORK}
In this section, we briefly discuss the basic idea regarding the construction of recurrence networks.  
More details can be found in recent reviews on the topic \cite {don2,don3}.
Recurrence is a fundamental property of every bounded dynamical system by which a trajectory tends to revisit 
a certain region of the phase space over a time interval. This basic concept has been utilized to 
develop a visualization tool called the recurrence plot (RP) for the analysis of dynamical systems 
\cite {eck}. A RP represents all recurrences in the form of a binary matrix $\mathcal R$ 
where $R_{ij}$ = 1 if the state $\vec {x_j}$ is a neighbour of $\vec {x_i}$ in phase space and 
$R_{ij} = 0$, otherwise. The neighbourhood is defined through a certain recurrence threshold $\epsilon$. 
In the most general definition, the discretely sampled scalar time series 
$s(1), s(2), .....s(N_T)$ is embedded in $M$-dimensional space taking the time delay co-ordinates 
\cite {gra} using a suitable time delay $\tau$, where $N_T$ is the total number of points in the time 
series. The procedure creates delay vectors in the embedded space of dimension $M$ given by
\begin{equation}
\vec {x_i} = [s(i),s(i+\tau),...... s(i+(M-1)\tau)]
  \label{eq:1}
\end{equation} 
There are a total number of $N = N_T - (M-1)\tau$ state vectors in the reconstructed space representing 
the attractor. Any point $\jmath$ on the attractor is considered to be in the neighbourhood of a reference 
point $\imath$ if their distance in the $M$-dimensional space is less than the threshold $\epsilon$. 
Thus we have 
\begin{equation}
R_{ij} = H (\epsilon - ||\vec {x_i} - \vec {x_j}||)
  \label{eq:2}
\end{equation}
where $H$ is the Heaviside function and $||..||$ is a suitable norm. In this paper, we use the 
Euclidean norm. The RP can only visually distinguish between different qualitative features of dynamics. 
This tool has become more popular with the introduction of the recurrence quantification analysis (RQA) 
\cite {mar2} using the measures derived from the RP. It has found numerous applications \cite {giu,fac,lit} 
and even dynamical invariants like correlation dimension $D_2$ and correlation entropy $K_2$ can be 
evaluated efficiently using RQA \cite {thi}.

The importance of the $\epsilon$ - RN (which, from now on, we simply call RN) is that its generation is 
closely associated with the RP. In fact, the adjacency matrix $\mathcal A$ for the unweighted RN can be 
obtained by removing the identity matrix from the recurrence matrix:
\begin{equation}
A_{ij} = R_{ij} - \delta_{ij}
  \label{eq:3}
\end{equation}
where $\delta_{ij}$ is the Kronecker delta. Note that, once the adjacency matrix is defined, the time 
series has been converted into a complex network. Each point on the embedded attractor is taken as a 
node in the RN and a node $\imath$ is connected to another node $\jmath$  if the distance $d_{ij}$ between the  
corresponding points on the embedded attractor is $\leq \epsilon$. Thus 
the adjacency matrix $\mathcal A$ is a binary symmetric matrix with elements 
$A_{ij} = 1$  if  ${d_{ij} \leq \epsilon}$ and $0$ otherwise. 
Note that, in contrast to the RP measures which consider the temporal properties of the trajectory 
points, RN analysis  quantifies the geometrical properties of 
the underlying attractor  and hence can give useful information regarding the structure of the attractor. 

Even though the method for the generation of the RN appears to be simple, it has several ambiguities 
associated with it \cite {don4}.  
How do we ensure that the RN captures the structural  characteristics of 
the underlying attractor? The answer lies in the proper choice of the parameters involved in the 
construction of the network. For the 
RN, the key parameters are $\epsilon$ and $M$. If $\epsilon$ is large, specific small scale properties 
of the attractor cannot be revealed and if $\epsilon$ is too small, the network breaks into 
dissuaded nodes due to lack of connections. Many authors \cite {dong1,gao2,sch,dong2} have discussed 
this issue in detail and have given some guidelines for the choice of $\epsilon$. But for arbitrary 
size of the attractor, the choice of $\epsilon$ still remains subjective. Similarly, the specific 
feature of RN generation is embedding and in this context the choice of $M$ has not been discussed much in the 
literature as it is commonly believed that $M$ should be sufficiently high for the attractor to be 
fully resolved. We show that the choice of $\epsilon$ is closely related to that of $M$ and we present 
a scheme for the choice of $\epsilon$ and $M$ that gives a uniform critical range of $\epsilon$ for all time series 
for a given $M$. To validate the wide range of applicability of the scheme, we show 
results from several low dimensional chaotic attractors as well as random data.

\section{\label{sec:level1}CHOICE OF PARAMETERS FOR NETWORK CONSTRUCTION}
There are four parameters associated with the RN generation, which are the time delay 
$\tau$, $\epsilon$, $M$ and $N$, the number of nodes. Note that $N < N_T$, the total number of points 
in the time series and the difference depends on $M$ and $\tau$. The value of $N_T$ can be adjusted to get 
the required number of $N$ for the computation. For the choice of $\tau$ we stick to the most commonly  
used criteria, namely, the first minimum of the autocorrelation function. The value of $\tau$ is related to 
the time step $\Delta t$ used for the generation of the time series. For the sake of uniformity, we 
use $\Delta t = 0.05$ to generate the time series from all the continuous time systems presented 
here. We have removed the first $10000$ values as transients in all cases. 

In all our numerical computations we use the value of $N$ in the range $2000$ to $10000$. The lower 
limit is set because, if the number of data points in the time series is too small,  
the basic structure of the embedded attractor may not have evolved completely. The upper limit is set mainly 
due to the fact that the computations become increasingly difficult due to high memory requirement for  
$N > 10000$. However, we find that $N < 10000$ is sufficient to get reasonable results from 
low dimensional chaotic systems. Moreover, for many real world applications of RN analysis, one  
has to  confine to this range of $N$ very often.

We next consider the choice of the crucial parameter, namely, the critical threshold, $\epsilon_c$. 
There are already a number of criteria suggested for choosing $\epsilon_c$. For example, Gao and 
Jin \cite {gao2} gives a heuristic criterion that selects $\epsilon_c$ as the value at which the 
link density becomes maximum when plotted as a function of $\epsilon$. But this has the drawback 
that small changes in $\epsilon$ induce large changes in the network measures as indicated by 
Donner et al. \cite {don4}. Recently, another criterion has been suggested based on analytic 
methods \cite {dong2} while an adaptive selection of threshold has been proposed by Eroglu et al. 
for specific time series \cite {ero1}. The last one is especially important as it chooses the 
critical threshold based on the theory of random graphs, where the second smallest eigen value 
$\lambda_2$ of the Laplacian matrix $\mathcal L$ crosses zero when plotted as a function of $\epsilon$ as the 
network becomes fully connected \cite {boc}, where $\mathcal L = \mathcal D - \mathcal A$, the 
difference between diagonal degree matrix and adjacency matrix.  
We will show below that this criterion comes  very close to the empirical criterion used by us. 

In all the above works so far considered, the size of the attractor after embedding is arbitrary 
so that the value of the threshold will be different for different attractors. Our primary motivation 
in the present work is to look for a scheme that can give approximately identical value for the 
critical threshold for different time series. We consider this to be an important step forward as 
it will lead to a uniform framework for the RN analysis which may be more useful for application to 
practical time series, as shown below.  

To overcome the problem of arbitrary size of the attractor, we  transform 
the time series into a \emph {uniform deviate}. For this, we first rescale the time series into the 
unit interval $[0,1]$. We then take each value $y_i$ in the time series and count how many values are 
less than or equal to $y_i$. Let this count be $n_i$. Then the uniform deviate time series $u_i$ is 
obtained as $u_i = {{n_i} \over N_T}$ where $N_T$ is the total number of points in the time series.   
The effect of uniform deviate transformation is shown in Fig.~\ref{f.1} for the Lorenz attractor, where the 
original time series $y(t)$ and the time series after uniform deviate $u(t)$ are shown along 
with the corresponding attractors after embedding. We have shown the importance of uniform 
deviate transformation in computing the conventional nonlinear measures like correlation 
dimension $D_2$ and entropy $K_2$ \cite {kph1, kph2}, especially from higher dimensional systems 
\cite {kph3}. It stretches the embedded attractor  uniformly in all directions  
without changing any of the dynamical invariants of the 
attractor, minimises the edge effect 
and provides improved scaling region and better convergence with data points. 

\begin{figure}
\includegraphics[width=0.96\columnwidth]{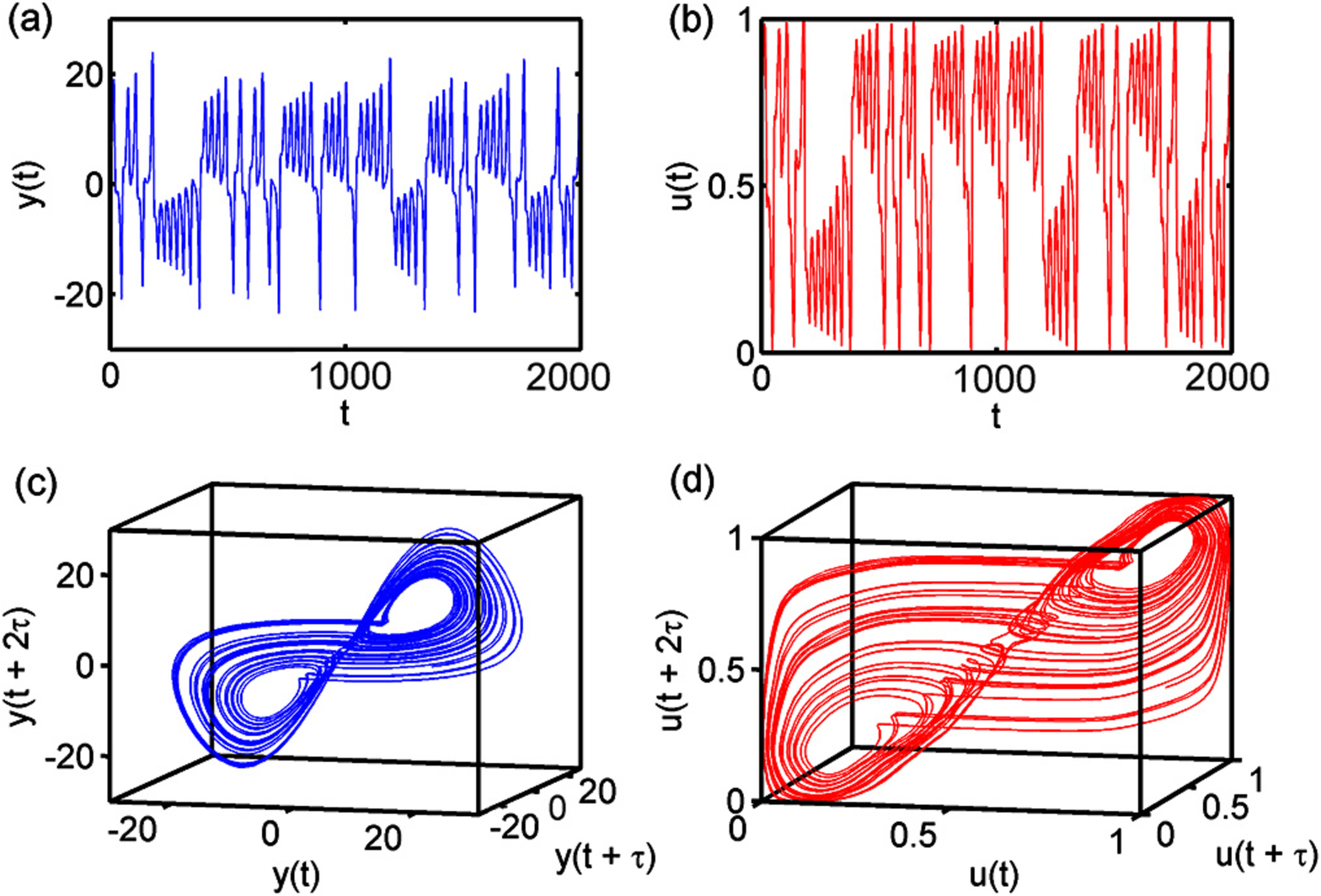}%
\caption{\label{f.1}(Color online) Top panel shows the original time series $y(t)$ from the Lorenz 
attractor on the left and the time series $u(t)$ after uniform deviate transformation on the right. 
The bottom panel shows the corresponding attractors after embedding.} 
\label{f.1}
\end{figure}

The primary criterion that we use for the selection of $\epsilon_c$ in our scheme is that 
the resulting RN has to remain mostly as ``one single cluster''. This is checked empirically 
by changing $\epsilon$. Note that this criterion is analogous to the one suggested by Donges et al. 
\cite {dong2} where the authors propose to choose  the percolation threshold above which the 
giant component of the RN appears. 
For the selection of $\epsilon_c$, we use a more practical approach rather than the rigorous criterion as 
adopted by the previously mentioned authors and we show the advantages of our approach in the sections below. 
A formal method to select $\epsilon_c$ is to choose the value of $\epsilon$ when the network just becomes fully 
connected. This is done by computing the second smallest eigen value $\lambda_2$ of the Laplacian 
matrix as a function of $\epsilon$. From the theory of random graphs, it is well known \cite {boc} 
that as $\lambda_2$ crosses zero from negative, the network becomes fully connected, with no dissuaded nodes. 
In  Fig.~\ref{f.2}, we show the variation of $\lambda_2$ with $\epsilon$ for RNs constructed from 
Lorenz and random time series, for $N = 2000$ and $5000$, with $M = 3$. It is evident that $\lambda_2$ for random 
network becomes positive slightly earlier compared to Lorenz in both cases. As $N$ increases, there is 
also a slight shift towards lower $\epsilon$. The network for random becomes fully connected for 
$\epsilon = 0.09$ while for Lorenz, the value is $0.13$. We have repeated the computation for other 
standard chaotic attractors and found that the value of $\epsilon$ where $\lambda_2$ becomes 
positive varies slightly for different systems. However, we find that 
there is a small range of $\epsilon$, from $0.1$ to $0.13$, where the network is \emph {almost}  
fully connected for all systems with the appearance of a \emph {giant cluster} and very few 
$(< 1\%)$ dissuaded nodes. This range is found to be common for all the 
systems we have analysed for a fixed $M$. Thus, the primary criterion provides a small uniform 
range of $\epsilon$ for many standard chaotic systems. 

\begin{figure}
\includegraphics[width=0.96\columnwidth]{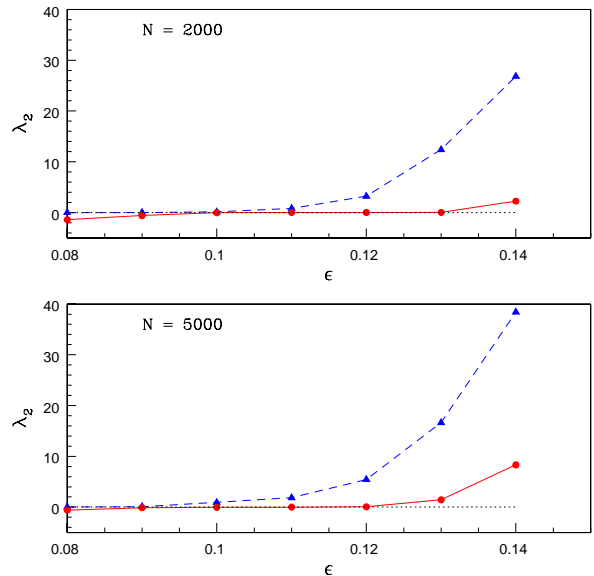}%
\caption{\label{f.2}(Color online) Variation of the second smallest eigen value $\lambda_2$ of the Laplacian 
matrix as a function of $\epsilon$ for the RNs from Lorenz (filled circles connected by red solid line) and random 
(filled triangles connected by blue dashed line) time series. 
Results are shown for RNs with $N = 2000$ and $5000$, constructed with $M = 3$. 
The dotted line is a reference for zero.} 
\label{f.2}
\end{figure}

In order to ensure that the RN is a true representation of the 
chaotic attractor, we apply an additional constraint that RN measures from standard chaotic 
time series are significantly different from that of a random time series. Though this 
condition appears to be subjective, we show below that it gives consistent results for all 
the standard chaotic systems considered in this work. 
For this, we compute the three important network measures, LD, CC 
and CPL for the RN from some standard chaotic times series as a function of $\epsilon$ and 
compare them with the corresponding measures from the RN of random time series. The equations for 
computing these measures are discussed in detail in the next section. 
We then find that the measure CPL is a good candidate to apply this additional constraint. 
This is shown in  Fig.~\ref{f.3} 
for two standard chaotic time series. For $\epsilon$ below $0.1$, there are multiple 
disconnected networks with no giant cluster and the CPL computed is for the largest component. 
It is evident from the figure that there is a small range of $\epsilon$,   
say $\Delta \epsilon$, (marked by the two vertical dashed lines) where the difference in the value 
of CPL of the RN from chaotic time series and that of random is maximum and 
as $\epsilon$ increases  above this range, the CPL for RN from all chaotic time series 
approaches that of random time series.  
More importantly, this range, which we call 
the \emph {critical range}, is found to be approximately identical for all the systems we have  
analysed and coincides with the common range found above corresponding to the emergence of 
the giant component in the RN.  
However, it should be emphasized that this range, $\Delta \epsilon$, is an empirical result 
and hence it is difficult to set 
any specific criterion for the upper bound either in terms of  $M$ or $N$.  
We choose the minimum value of this range as the critical threshold  
$\epsilon_c$, for the construction of RN which we believe will capture the characteristic 
properties of the attractor. Nevertheless, we have checked and confirmed that any $\epsilon$ within  
$\Delta \epsilon$ does not make any qualitative change in the degree distribution of the RN and the 
related network measures to be presented below. Note that we do not follow the condition 
$\lambda_2 > 0$ strictly as chosen by Eroglu et al. \cite {ero1}, as this makes $\epsilon_c$ 
slightly different for different RNs. We have constructed the 
RN using the Gephi software \emph{(https://gephi.org/)} taking time series from several standard chaotic 
attractors  for $\epsilon$ ranging from $0.05$ to $0.25$, taking $M = 3$. We find that there is no giant 
cluster for $\epsilon < 0.1$ while the network becomes over connected for $\epsilon > 0.15$, in all cases.  

\begin{figure}
\includegraphics[width=0.96\columnwidth]{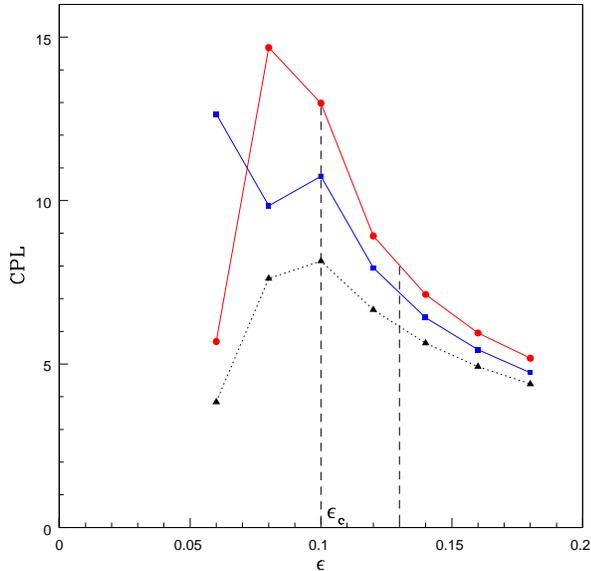}%
\caption{\label{f.3}(Color online) Characteristic path length as a function of 
threshold for RNs generated from Lorenz (filled circles connected by red solid line at the top), 
R\"ossler (filled squares connected by blue solid line in the middle) and 
random (filled triangles connected by dotted line) time series 
with $N = 2000$ and $M = 3$. The two vertical dashed lines indicate the critical range of 
$\epsilon$ where CPL from chaotic time series differs from that of random time series. The 
critical threshold $\epsilon_c$ is the minimum of this range. Below $\epsilon_c$, there is 
no giant cluster.} 
\label{f.3}
\end{figure}

We now consider how the value of $\epsilon_c$ changes with $M$ and $N$. To study this, we generate RNs 
from a number of low dimensional chaotic systems, both discrete and continuous, by varying $N$ 
from $2000$ to $10000$ and $M$ from $2$ to $5$. For each $N$ and $M$, we scan a range of 
$\epsilon$ values between $0.02$ to $0.20$. The results from this detailed numerical analysis 
are compiled in Fig.~\ref{f.4} separately for continuous and discrete systems. The variation of 
$<l>$ corresponding to $\Delta \epsilon$ as $N$ and $M$ changes is shown with that of 
random time series as reference. The left panel shows the dependence on $N$ and the right panel 
shows the dependence on $M$. To study the dependence on $N$, we use the natural dimension of 
the system, namely, $M=3$ for continuous systems (top) and $M=2$ for discrete systems (bottom). 
Note that  $\Delta \epsilon$  for $M=2$ has been shifted to $0.06 - 0.08$. Moreover, 
in both cases, ($M=2$ and $3$), there is only a small decrease in  
$\Delta \epsilon$ as $N$ is increased from $2000$ to $10000$. This means that one can effectively use 
the same $\epsilon$ for this whole range of $N$ values.

\begin{figure}
\includegraphics[width=0.96\columnwidth]{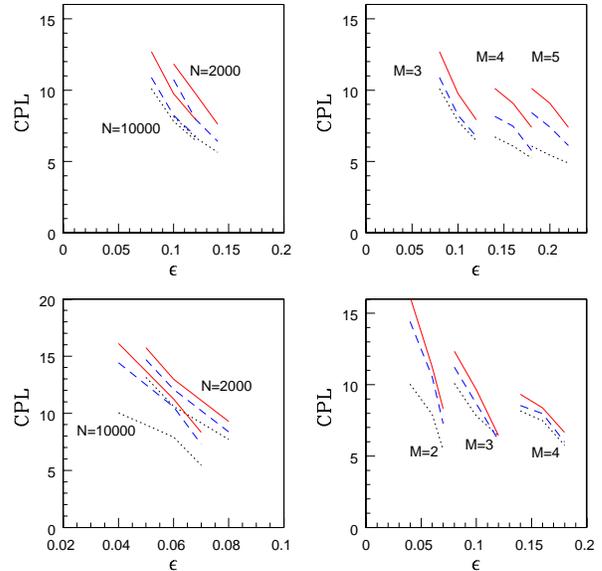}%
\caption{\label{f.4}(Color online) The figure shows how the critical range $\Delta \epsilon$ 
(see text) varies with respect to  $N$ and $M$. The left panel shows  $\Delta \epsilon$ 
for two values of $N$ as indicated, where the top panel is for continuous systems with 
$M=3$ and the bottom panel for discrete systems with $M=2$. The right panel shows the 
variation of  $\Delta \epsilon$ with $M$ for $N$ fixed at $10000$. In the top 
panel, the red solid line is for Lorenz attractor and the blue dashed line is for R\"ossler attractor, 
while in the bottom panel, the same is for Lozi and Henon attractors respectively. In 
all figures, the dotted line is for random time series.} 
\label{f.4}
\end{figure}

In the right panel, we show the effect of increasing the embedding dimension from the natural 
dimension of the system, with $N$ fixed as $10000$. Note that the value of $\epsilon_c$ clearly shifts 
with $M$ which implies that each $M$ requires a corresponding $\epsilon_c$ for the generation of 
the RN. But the more interesting result is that, for all the systems that we have analysed, we are 
able to find   
approximately the same critical range $\Delta \epsilon$ corresponding to each $M$. This range is 
given in Table 1 where, $\epsilon_c$ in the third column is the critical threshold  
used by us corresponding to each $M$ for all the further computations in this paper. Thus, 
even though our criterion for the selection of $\epsilon_c$ is not completely novel, we are 
able to provide a certain level of non-subjectivity in its choice which, we hope, will make 
the application of RN analysis more effective.

It may be noted that the above choice of the critical range of $\epsilon$ is, in fact, 
analogous to and motivated by the selection of a \emph {scaling region} in the 
conventional nonlinear time series analysis for deriving dynamical invariants like 
$D_2$. We have already shown that the scaling region for many low dimensional 
chaotic systems can be selected algorithmically for a non-subjective computation of 
$D_2$ and $K_2$ \cite {kph1,kph4}. It is well known that the choice of the 
scaling region critically depends on the embedding dimension $M$, while the change is 
not much if the number of data points changes from $2000$ to $10000$. In this way, 
the above result of getting an approximately same range of $\Delta \epsilon$ for 
different chaotic systems is not very surprising. 

\begin{table}[h]
\centering
\begin{tabular}{|l|c|c|}
\hline
M & \emph{Critical Range of $\epsilon$} & $\epsilon_c$ \\
\hline

2 & 0.06 - 0.08 & 0.06\\

& & \\

3 & 0.10 - 0.13 & 0.10\\

& & \\

4 & 0.14 - 0.18 & 0.14\\

& & \\

5 & 0.18 - 0.22 & 0.18\\

\hline
\end{tabular}
\caption{Critical range of $\epsilon$ obtained empirically for each value of $M$}
\label{tab:1}
\end{table}

We now give a simple mathematical explanation for the observed numerical results regarding the 
relation between $\epsilon$ and $M$. Consider a random distribution of $N$ points in $M$ 
dimension. After uniform deviate transformation, the volume of the embedding space is 
unity and the average density of points $< \rho > = N$. The average separation between two 
points along any direction is 
\begin{equation}
<d>  \sim ({{1} \over {N}})^{1/M} \sim ({{1} \over {< \rho >}})^{1/M}
  \label{eq:4}
\end{equation}
This gives the critical value of distance  for a given $M$ below which 
the degree of a node tends to zero on the average. Hence  
$\epsilon_c$ for the random RN for each $M$ must be sufficiently 
greater than $<d>$.  

To get more insight on the result given in Table 1 and know how $\epsilon_c$ varies for 
higher $M$ values, we consider the \emph {limiting value} of $\epsilon$, say $\epsilon_f$, 
at which the RN is fully connected. That is, every node is connected to every other node 
with the degree of each node $(N-1)$ and the LD reaches its maximum possible value $1$. 
Since we consider a uniform deviate, the size of the attractor is unity. For $M = 2$, 
the value of $\epsilon$ at which this happens is the diagonal length of the square, 
that is, $\epsilon_f(2) \equiv \sqrt 2$. For $M = 3$, $\epsilon_f(3)$ increases to 
$\sqrt 3$ and one can easily show that, in general, $\epsilon_f(M) \equiv \sqrt M$. 
Note that this result is independent of $N$. This also implies that the difference 
between successive $\epsilon_f$, $\epsilon_f(M) - \epsilon_f(M-1)$, slowly decreases 
with $M$. Since the LD corresponding to $\epsilon_c$ is effectively a fraction of the 
total LD, one expects roughly the same dependence for $\epsilon_c$ on $M$, that is  
$\epsilon_c \propto \sqrt M$. We have numerically verified this for the random RN for 
$M$ upto $7$.  This means that, though the difference between 
successive $\epsilon_c$ appears to be almost a constant for small $M$ as given in 
Table 1, this difference decreases slowly as $M$ increases.

As a final test to validate our choice of $\epsilon_c$, we undertake a 
counter check by computing $D_2$ of some standard chaotic attractors using the 
RP (which is equivalent to the 
adjacency matrix) corresponding to  $\epsilon_c$. The method proposed by 
Thiel et al. \cite {thi}  is used for this purpose.
This method uses the cumulative probability distribution 
$p^c(l)$ of the diagonal lines in the RP, corresponding to  two 
different thresholds $\epsilon$ and $\epsilon + \Delta \epsilon$ using the relation: 
\begin{equation}
D_2(\epsilon) = {{log [{{p^c(\epsilon,l)} \over {p^c(\epsilon+\Delta \epsilon,l)}}]} \over {log [{{\epsilon} \over {\epsilon+\Delta \epsilon}}]}}
  \label{eq:5}
\end{equation}
We have found that the $D_2$ values obtained in all cases are closer to the standard 
values for $\epsilon = \epsilon_c$. This implies that the geometric complexity of the 
attractor is truly reflected in the RN constructed by our scheme.
     
\section{\label{sec:level1}MEASURES FROM RECURRENCE NETWORK} 
We now compute the important 
network measures from the RN of several low dimensional chaotic attractors, including 
discrete systems (maps) in two dimensions and continuous systems (flows) in three 
dimensions. For all systems, $M$ is varied from 2 to 5 using the $\epsilon_c$ corresponding to 
each $M$ with $N$  varied from $2000$ to $10000$. 

\subsection{Characteristic Path Length, Clustering Coefficient and Link Density}
We first compute the CPL, CC and LD from the RN and study their dependence on $N$ 
and $M$. The equations for computing these measures 
have been discussed in detail in the literature \cite {new1,alb1}.
The CPL is given by the equation
\begin{equation}
<l> = {{1} \over {N(N-1)}} \sum_{i,j}^N l_s
  \label{eq:6}
\end{equation}
where $l_s$ is the shortest path length for all pair of nodes $(\imath,\jmath)$ in the network. 
The maximum value of $l_s$ is taken as the diameter of the network, $l_D$. 
If $k_i$ is the degree of the $\imath^{th}$ node, then
\begin{equation}
LD = {{1} \over {N(N-1)}} \sum_{i}^N k_i
  \label{eq:7}
\end{equation}  
The CC of the network is defined through a local clustering index $c_v$. Its value is obtained 
by counting the actual number of edges in a sub graph with respect to node $v$ as reference to 
the maximum possible edges in the sub graph:
\begin{equation}
c_v = {{\sum_{i,j} A_{vi}A_{ij}A_{jv}} \over {k_v(k_v - 1)}}
  \label{eq:8}
\end{equation}  
The average value of $c_v$ is taken as the CC of the whole network:
\begin{equation}
CC = {{1} \over {N}} \sum_{v} c_v
  \label{eq:9}
\end{equation}  
In Fig.~\ref{f.5} top  panel, we show the variation 
of $<l>$ with $N$ for two standard chaotic systems for $M = 3$. RN from random 
time series is also added for comparison. Note that in all cases, $<l>$ initially 
decreases with $N$, but saturates as $N \rightarrow 10000$, with the value of $<l>$ 
for RN from random time series always less compared to that from chaotic systems. 
The variation of $<l>$ with $N$ can be understood from the degree distribution of the RN 
discussed in detail in the next section. We find that as $N$ increases, the average 
degree of the nodes $<k>$ also increases correspondingly. Typically, as $N$ increases to 
$2N$, $<k>$ shifts approximately to $<2k>$, reducing $<l>$. In the bottom panel, we show 
the variation of CC with $N$ which is found to be approximately constant for all systems 
for the range of $N$ used.  

\begin{figure}
\includegraphics[width=0.96\columnwidth]{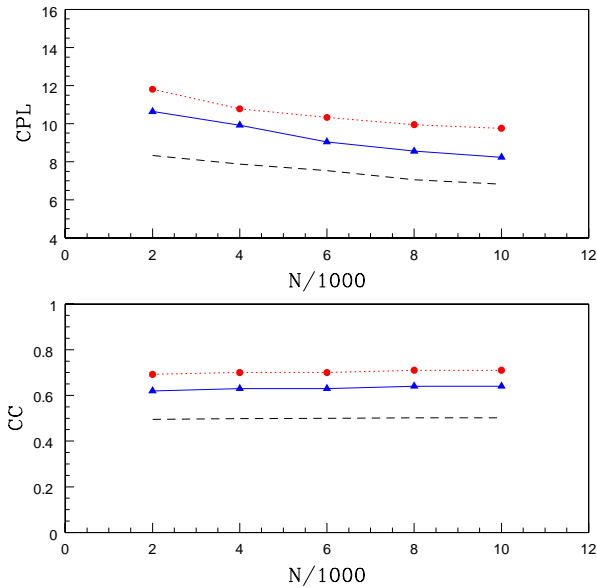}%
\caption{\label{f.5}(Color online)Variation of CPL (top panel) and CC (bottom panel) with 
$N$ for RNs generated from Lorenz (filled circles connected dotted line), 
R\"ossler (filled triangles 
connected by solid line) and random (dashed line) time series. The value of $M$ is fixed as $3$.} 
\label{f.5}
\end{figure}

\begin{figure}
\includegraphics[width=0.96\columnwidth]{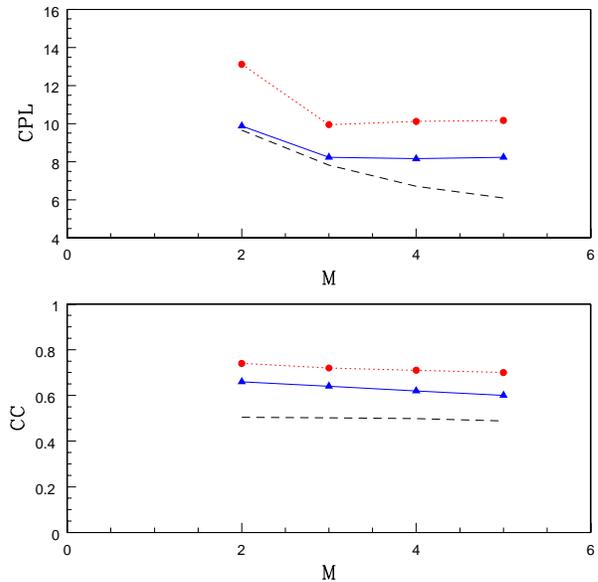}%
\caption{\label{f.6}(Color online)Same as the previous figure, but the variation is 
with respect to $M$ for a fixed $N = 10000$. Note that CPL saturates at the natural 
dimension of the system.} 
\label{f.6}
\end{figure}

We have also studied numerically the variation of $<l>$ and CC with $M$ by fixing 
$N = 10000$ and the results for the same two systems are shown in Fig.~\ref{f.6}. 
While CC remains constant, the variation of $<l>$ is more interesting, showing 
saturation for $M$ equal to or greater than the actual dimension of the system. 
We will show below that the same is true for degree distribution as well, leading to 
an important practical application of network measures. 

\subsection{Degree Distribution}
We now consider the most important measure of a network, namely, the degree 
distribution $P(k)$ versus $k$.  
In Fig.~\ref{f.7}, we show the degree distribution of the RN from 
Lorenz and R\"ossler attractor time series for $N=5000$ and $10000$ by using $M=3$ and 
$\epsilon = 0.1$. Note that the error bar is estimated from counting statistics 
resulting from the finiteness in the number of nodes. The statistical error associated 
with any counting of $n(k)$ is $\sqrt {n(k)}$. If $n(k) \rightarrow 0$, one typically 
takes the error to be normalised as 1. Thus, the error associated with $P(k)$ is 
typically ${{\sqrt {n(k)}} \over {N}}$ and becomes $1/N$ as $n(k) \rightarrow 0$. 
It is evident from the figure that as $N$ is doubled, the degree $k$ of each node gets 
approximately doubled resulting in a shift along the X-axis and the range 
$[k_{min}, k_{max}]$ of $k$ values is shifted approximately to twice the range. 
Correspondingly, the $P(k)$ values are reduced since the area under the distribution curve 
is constant. Assuming that the $k$ values are continuous within the range 
$[k_{min}, k_{max}]$, we can write 
\begin{equation}
\sum_{k} \Delta k P(k) = 1
  \label{eq:10}
\end{equation}
where $\Delta k (\equiv 1)$ is the difference between successive $k$ values. 

Changing $k$ to ${k^{'}} \equiv {{k} \over {N}}$, $\Delta {k^{'}} = {{\Delta k} \over {N}}$ and 
$P(k)$ changes to $P(k^{'})$ so that 
\begin{equation}
\sum_{k^{'}} \Delta k^{'} P(k^{'}) = 1
  \label{eq:11}
\end{equation}
Substituting for $\Delta k$ in Eq.(10), we get 
\begin{equation}
P(k^{'}) = N P(k) \equiv n(k)
  \label{eq:12}
\end{equation}
It is convenient to represent the degree distribution in the rescaled variables as shown 
in Fig.~\ref{f.8}. Note that the degree distribution for the two $N$ values have now become 
identical and can be considered as almost \emph {stationary} in the rescaled variable $k/N$ 
apart from small statistical fluctuations. To capture the real trend in the distribution, we show  
in Fig.~\ref{f.9} the same distributions shown in the previous figure without error bar and the 
values connected by line. 

One expects the scale invariance for the RN from a random time series whose degree distribution is Poissonian  
which can be approximated as Gaussian for large $N$. This is because, the degree of every node 
increases by an average value as $N$ increases and the degree distribution appears identical in the 
rescaled variable $k/N$ as can be seen from Fig.~\ref{f.10}.
Here we find that an arbitrary degree distribution from the RN of a 
chaotic attractor also shows this property. A possible explanation, for attractors whose measure is  
continuous is that, as the dynamical 
system evolves, the structure of the attractor also evolves in such a way that the probability 
density over the attractor is preserved once the basic structure of the attractor is formed. 
The degree $k_i$ of a reference node $\imath$ represents the local 
connectivity of the RN and it corresponds to the local phase space density around the reference 
point in the chaotic attractor from which the RN is constructed. It is well known that the local 
phase space density of the chaotic attractor is preserved in the RN \cite {don2}. 
Considering an 
infinitesimal hyper volume $V_M(\epsilon)$ in $M$-dimension with radius $\epsilon$ about a 
reference point $\vec r_i$ in phase space, one can write \cite {don2}:
\begin{equation}
{{1} \over {N}} (k_i(\epsilon)+1) \approx \int_{V_M(\epsilon)} p(\vec r)d^M\vec r \approx V_M(\epsilon) p(\vec r_i)
  \label{eq:13}
\end{equation}
where $p(\vec r_i)$ is the invariant density around $\vec r_i$. Note that in LHS, $1$ is added to 
include the reference node (self loop). This gives a relation between the local 
measure in an attractor and that of a RN:
\begin{equation}
p(\vec r_i) = \lim_{\epsilon \rightarrow 0} \lim_{N \rightarrow \infty} {{(k_i + 1)} \over {V_M(\epsilon) N}}
  \label{eq:14}
\end{equation}
The above equation tells us that for the RN constructed with $\epsilon_c$, the local 
probability density around a point on the attractor gets mapped to the degree of the 
corresponding node in the constructed RN. Since every point on the attractor is converted to a 
node on the RN, points with the same probability density will correspond to nodes with the 
same degree. The degree distribution tells how many nodes have a given degree in the RN. 
This is equivalent to finding how many local regions on the entire attractor have the same 
probability density. As we change from the phase space domain of the attractor to the 
domain of the network, the degree distribution represents the global statistical measure of 
the probability density variations over the entire attractor.  
Thus, it seems natural that the degree distribution of the RN from any chaotic 
attractor shows the scale invariance. The small deviations in the degree distribution as $N$ 
increases is the result of the corresponding small fluctuations in the probability density. 
Also, the range of $k$ values in the RN is a measure of the range of variation of $p(\vec r)$ 
over the attractor. However, a direct relation connecting the probability distribution 
over the attractor and the degree distribution of the RN seems to be highly nontrivial 
owing to the fractal geometry of the attractor.

From the above discussion, it becomes clear that a peak at high $k$ value near $k_{max}$ 
in the degree distribution implies large number of relatively dense regions of high probability over 
the entire attractor. Many peaks in the degree distribution are indicative of large local 
density fluctuations over the attractor. For example, from  Fig.~\ref{f.8} and Fig.~\ref{f.9}, 
the fluctuation is much large for the RN from the Lorenz attractor compared to that of the 
R\"ossler attractor, though both have approximately the same range of 
$[k_{min}, k_{max}]$. In this sense, one can say that 
the degree distribution is a characteristic measure of the structural complexity  
of an attractor. Note that this idea has been pointed out by many authors before \cite {don2,zou2}. 
Once the basic structure of the attractor is formed, a further increase in 
the number of nodes does not change the degree distribution qualitatively. In other words, 
RN analysis appears to be a useful tool to get meaningful results regarding structural and 
topological properties of the attractors with less number of data points. We now show that 
there is a part in the degree distribution that corresponds to the Poisson distribution 
where, the $k$ values occur more by chance than by choice. 

\begin{figure}
\includegraphics[width=0.96\columnwidth]{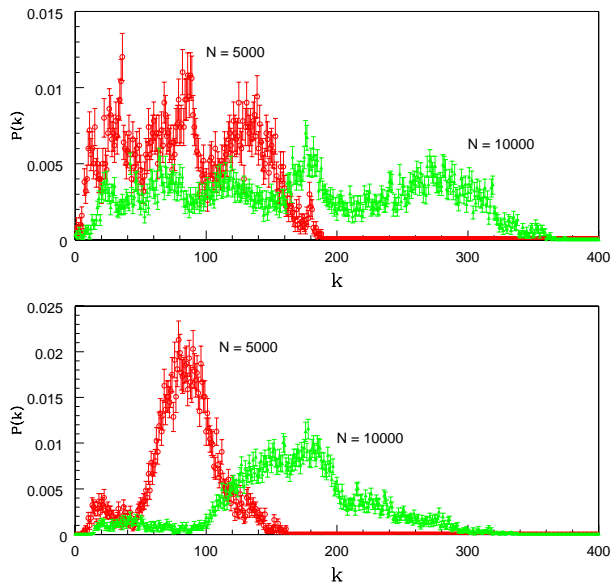}%
\caption{\label{f.7}(Color online) Degree distribution of the RN generated from the Lorenz attractor  
(top panel) for $N=5000$ (open red circles) and $N=10000$ (filled green triangles appearing in light gray 
shade in print), as indicated. Corresponding results for 
R\"ossler attractor are shown in the bottom panel. In both cases, we use $M=3$ and $\epsilon_c=0.1$.} 
\label{f.7}
\end{figure}

For a random time series embedded in $M$-dimensional space, after being converted into 
uniform deviate, the average density of points   
$< \rho > = N$. Hence the average number of points inside a $M$-dimensional sphere 
of radius $\epsilon$ is $k_{ran} = N V_M$, where $V_M$ is the volume of the sphere. When the 
time series is converted into a RN, the condition for two nodes to be connected is that the 
distance is $< \epsilon$. In other words, for random RN, typically a node is connected to 
$k_{ran}$ other nodes. Or, most nodes will have degree $k_{ran}$ and the degree distribution 
tends to be a Poissonian around this value. 

\begin{figure}
\includegraphics[width=0.96\columnwidth]{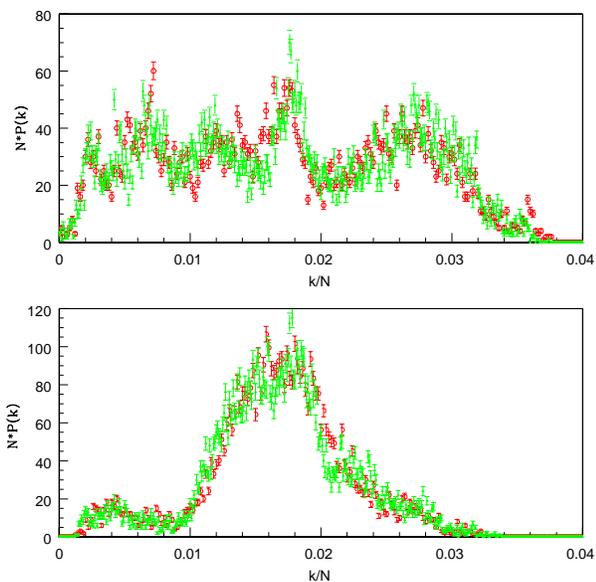}%
\caption{\label{f.8}(Color online) Rescaled distributions (see text) of the Lorenz (top panel) and 
R\"ossler (bottom panel) attractors 
shown in the previous figure for $N = 5000$ (open red circles) and $N = 10000$ (filled green 
triangles appearing in light gray shade in print).  
Note that, after rescaling, the distributions for the two $N$ 
values become statistically identical in both cases.} 
\label{f.8}
\end{figure}

\begin{figure}
\includegraphics[width=0.96\columnwidth]{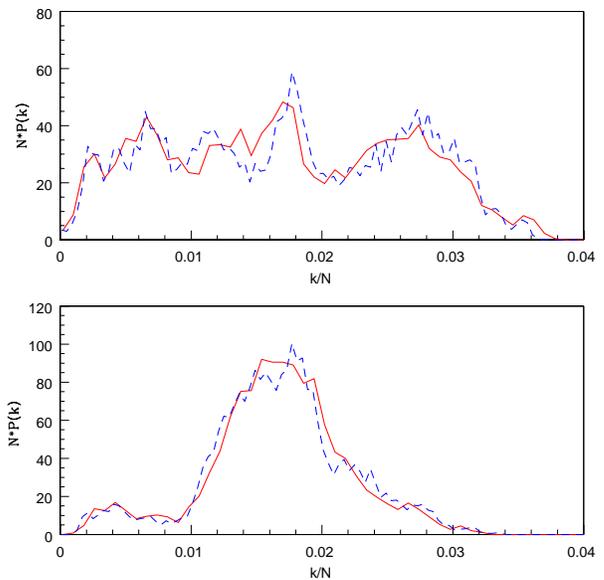}%
\caption{\label{f.9}(Color online) The same distributions shown in the previous figure  
without error bar and connected by line, for clarity. The red solid line is for $N = 5000$ and 
the blue dashed line is for $N = 10000$.} 
\label{f.9}
\end{figure}

Now, for a non random time series, there will be significantly more nodes with degree greater 
than $k_{ran}$ and it is these nodes which describe the structure of the underlying attractor. 
Nodes with degree $ \sim k_{ran}$ occur more by chance association rather than the true 
description of the system. Thus, characteristic information regarding the system is given by 
the nodes with degree $> k_{ran}$ and hence the value of $k_{ran}$ should be small compared to 
the range of $k$ values in the degree distribution. 

Note that the position of $k_{ran}$ in the degree distribution depends on the choice of 
$\epsilon$ and $M$. One expects a small peak around $k_{ran}$ in all degree distributions 
which becomes less significant as $N$ increases. We now estimate the value of $k_{ran}$ for 
the choice of $\epsilon$ and $M$ that we use to compute the degree distributions of chaotic 
attractors. For $M = 3$, $V_M = {{4} \over {3}} \pi \epsilon^3$ and hence 
\begin{equation}    
 k_{ran} \approx N{{4} \over {3}} \pi \epsilon^3 
 \label{eq:15}
\end{equation}
With $\epsilon = 0.1$, we get ${{k_{ran}} \over {N}} \approx 0.004$, which is independent of 
$N$. This is sufficiently small compared to the rescaled ${{k} \over {N}}$ values as can 
be seen from  Fig.~\ref{f.8} and Fig.~\ref{f.9}, where ${{k_{max}} \over {N}} \approx 0.04$. Since 
${{k_{ran}} \over {N}} \propto \epsilon^3$ for $M = 3$, a small increase in $\epsilon$ can 
shift the Poisson range significantly to the right in the degree distribution. This shows 
the importance of choosing the correct $\epsilon_c$.

\begin{figure}
\includegraphics[width=0.96\columnwidth]{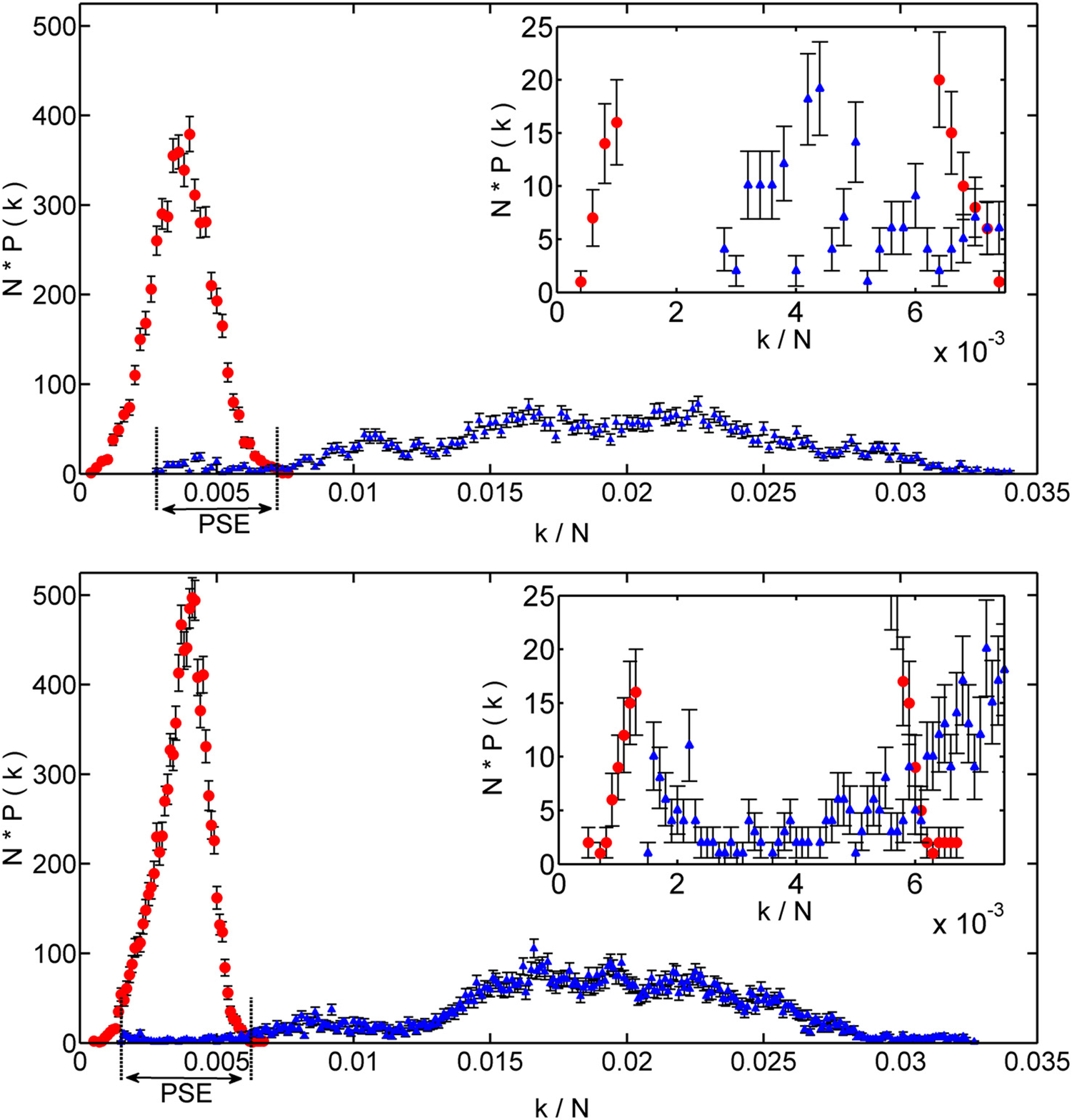}%
\caption{\label{f.10}(Color online) Degree distribution of the RN from the R\"ossler attractor time 
series (filled blue triangles)  in comparison with that of the random time series 
(filled red circles seen as Poisson distribution on the left).   
The top panel is for $N=5000$ and the bottom 
panel for $N=10000$, with $M=3$ and $\epsilon = 0.1$ in both cases. We show the rescaled 
distributions so that both panels appear identical. The two vertical lines indicate the 
Poisson statistics part of the distribution or Poisson statistic error (denoted PSE) 
which coincides with the degree distribution of 
the random time series. It is shown magnified in the inset.} 
\label{f.10}
\end{figure}

For the discrete systems with $M = 2$, we have
\begin{equation} 
 {{k_{ran}} \over {N}} \approx \pi \epsilon^2 
 \label{eq:16}
\end{equation}
Using the optimum value of $\epsilon = 0.06$ used for $M = 2$, we have 
${{k_{ran}} \over {N}} \approx 0.011$ which is $<< {{k_{max}} \over {N}}$, as will be 
shown below for discrete systems. Finally, for $M = 4$, we have a hyper cube of unit 
volume. The general formula for the volume of a Euclidean ball of radius $\epsilon$ in 
$M$-dimension (for even $M$) is:
\begin{equation}
V_M (\epsilon) = {{\pi^{M/2}} \over {(\frac {M}{2})!}} \epsilon^M
  \label{eq:17}
\end{equation}
For $M = 4$, $V_4 (\epsilon) = {{\pi^2} \over {2}} \epsilon^4$. For the threshold  
$\epsilon = 0.14$, we find ${{k_{ran}} \over {N}} \approx 0.00196$.

\begin{figure}
\includegraphics[width=0.96\columnwidth]{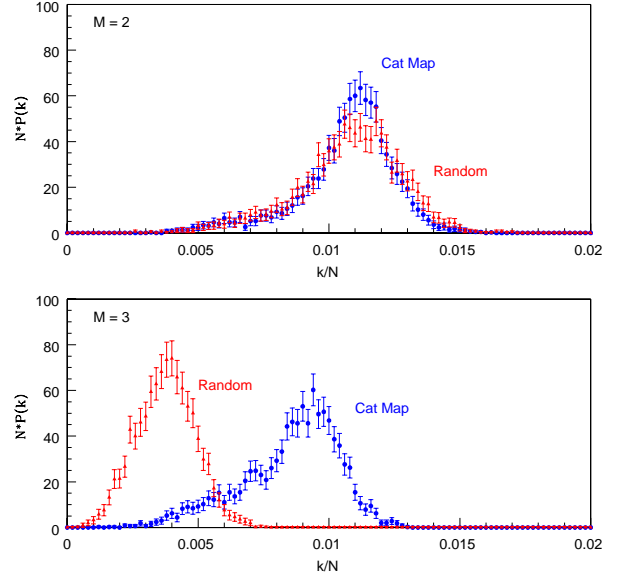}%
\caption{\label{f.11}(Color online) Degree distributions of the RNs generated from Cat Map 
(filled blue circles) and random (filled red triangles) 
time series for $N=5000$ for two values of $M$, as indicated. Note that, in going from $M=2$ to $M=3$, the 
distribution of the random network is shifted while that for the Cat Map remains stationary.} 
\label{f.11}
\end{figure}

The above results are explicitly shown in Fig.~\ref{f.10} and Fig.~\ref{f.11}.  
In Fig.~\ref{f.10}, we show the rescaled degree distributions of RNs 
from random and R\"ossler attractor time series plotted together for $N = 5000$ and 
$10000$. Note that the Poisson distribution part, shown by the two vertical lines, 
almost exactly coincides with the degree distribution of the random time series. 
This part is shown magnified in the inset. In Fig.~\ref{f.11}, we 
show the rescaled degree distributions of the Cat map and the random time series 
together for $M = 2$ (top panel) and for $M = 3$ (bottom panel). For $M = 2$, both 
the distributions are almost identical and peak exactly at $k_{ran} = 0.011$ in 
agreement with our calculations above. For $M = 3$, the peak for the random 
distribution is shifted to $0.004$ as expected, while that for Cat map remains  
almost unchanged and hence both the distributions can be easily differentiated. 

\begin{figure}
\includegraphics[width=0.96\columnwidth]{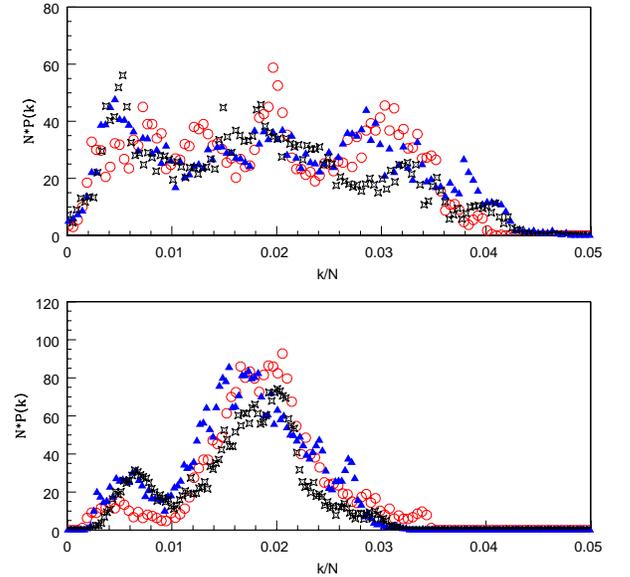}%
\caption{\label{f.12}(Color online) Rescaled degree distributions (without error bar for clarity) 
for the RNs from Lorenz (top panel) and R\"ossler (bottom panel) attractor time series for 
$M = 3$(open red circles), $M = 4$(filled blue triangles) and 
$M = 5$(black star like points).} 
\label{f.12}
\end{figure}

We have, so far, computed the degree distribution by taking the actual dimension of 
the attractor. However, in the analysis of the real world data, there is no 
\emph {a priori} information regarding the dimension of the system. Then one has to 
compute the network measures for different $M$ values and check for saturation.  
In Fig.~\ref{f.12}  we show the rescaled degree distributions of RNs from Lorenz and 
R\"ossler attractors for $M = 3,4,5$. For each $M$, the corresponding $\epsilon_c$ 
value found empirically, as given in Table 1, is used. We find that the degree distribution 
remains almost invariant (apart from small changes due to the effect of embedding), for 
$M$ values equal to or greater than the actual dimension of the system.
On the one hand, it is a counter check whether we are using the correct value of 
$\epsilon_c$ for each $M$ and on the other hand, the result tells us that the usual practice of 
using a high value of $M$ for real world data works for network measures as well provided the 
correct $\epsilon_c$ corresponding to each $M$ is used.   
However, a higher $M$ requires a correspondingly 
larger value of $N$. We find that it is sufficient to use $N < 10000$ and $M \leq 5$ 
for a proper characterization of low dimensional chaotic systems using RN measures. This 
also leads to a practical application of the proposed method discussed in more detail in the 
next section.

\begin{figure}
\includegraphics[width=0.96\columnwidth]{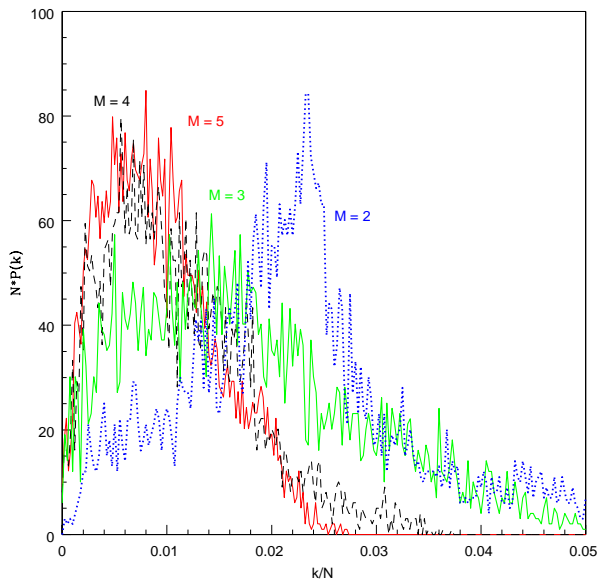}%
\caption{\label{f.13}(Color online) Degree distributions (without error bar) of the RN constructed 
from the time series 
of Chen hyperchaotic attractor for $M = 2$ (thick blue dotted line), $M = 3$ (thick green solid line), $M = 4$ 
(black dashed line) and $M = 5$ (thin red solid line). For each $M$, the corresponding value of $\epsilon_c$ 
given in Table 1 is used for constructing the RN with $N = 5000$.} 
\label{f.13}
\end{figure}

We next show that the present scheme is efficient not 
only for low dimensional chaotic attractors, but also for the analysis of high dimensional and hyperchaotic 
systems as well.  In Fig.~\ref{f.13}, we show the degree distributions 
of the RN constructed from the time series of Chen hyperchaotic flow \cite {chen} for $M$ varying from 
$2$ to $5$, with corresponding values of $\epsilon_c$. We generate the hyperchaotic time series with 
$5000$ data points using the standard set of parameters: $a = 35$, $b = 4.9$, $c = 25$, $d = 5$, $e = 35$ 
and $k = 100$. Note that the degree distributions for $M = 4$ and $5$ are almost identical while that for 
lower $M$ values deviate, since the attractor dimension is $> 3$. This is also a direct confirmation of 
our argument above that for a real world data whose dimension is unknown, one has to check for saturation of 
network measures by increasing the $M$ value. This is somewhat equivalent to finding the saturated 
$D_2$ value by increasing $M$ in the conventional nonlinear time series analysis.  

Another important outcome of our scheme is that we are able to compare the characteristic 
measures derived from the RN of different chaotic attractors since the analysis is done using 
identical threshold for a fixed $M$.  
For example, the degree distribution of the 
RN typically characterises the structure of the attractor, as discussed above.  
Hence, a visual inspection of the degree distribution can  
provide some qualitative information on the structural complexities of standard low dimensional 
chaotic attractors, as shown in Fig.~\ref{f.14} and Fig.~\ref{f.15}. The degree 
distribution in each case is the average from four RNs generated using different 
initial conditions for the attractor.  
From a comparison of the degree distributions in Fig.~\ref{f.14} one can infer that the 
Lorenz attractor is structurally more complex compared to the other three since the 
fluctuation in the probability density is maximum for it. More accurate results can be 
obtained from a quantitative analysis of the network measures. 

\begin{figure}
\includegraphics[width=0.96\columnwidth]{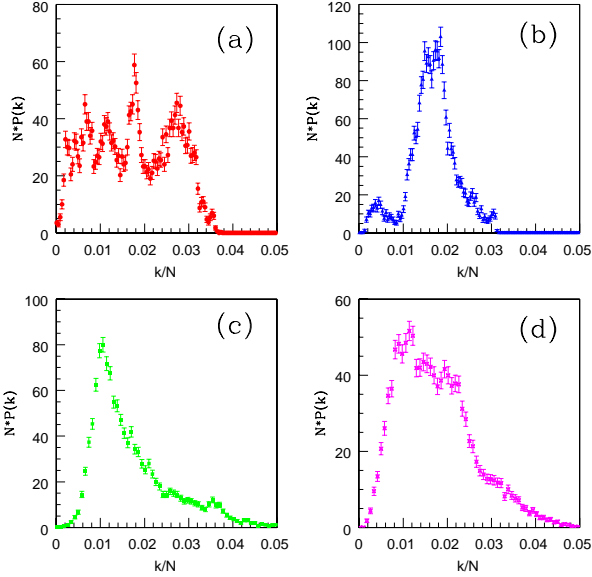}%
\caption{\label{f.14}(Color online) Characteristic degree distributions 
of four standard low dimensional chaotic attractors  for $M=3$. The systems are 
(a) Lorenz (b) R\"ossler (c) Duffing and (d) Ueda attractor. Standard parameter values 
given in \cite {spr} are used for the generation of time series from Duffing and Ueda 
attractors. The average degree distribution for RNs generated from four initial 
conditions is shown in all cases.} 
\label{f.14}
\end{figure}

\begin{figure}
\includegraphics[width=0.96\columnwidth]{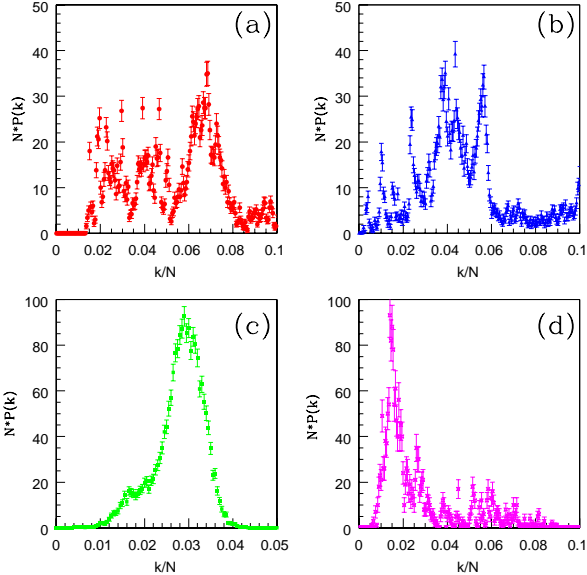}%
\caption{\label{f.15}(Color online) Same as the previous figure for four standard chaotic maps, 
namely, (a) Henon (b) Lozi (c) Cat Map and (d) Logistic Map.  
For the logistic map in the fully chaotic regime, we use $M=1$ and $\epsilon_c = 0.01$ which 
results in the high peak corresponding to that $\epsilon$ value. For all other cases, 
we use $M=2$.} 
\label{f.15}
\end{figure}

For the logistic map, we use the fully chaotic 
region with $M = 1$ and $\epsilon_c = 0.01$ and hence the very large peak in the 
degree distribution corresponding to that $\epsilon$ value is due to Poisson statistics. 
The logistic map requires a special mention. For an attractor in one 
dimension, the Poisson value of $k/N \equiv \epsilon_c$, the threshold itself. 
In other words, for a random distribution in the unit interval, the degree 
distribution is typically a Poissonian around degree $k/N \equiv \epsilon_c$ in our scheme.  
However, for the RN from the logistic attractor, depending on the probability density 
variations, the number of degrees of any node can be $>> \epsilon_c$ or 
$<< \epsilon_c$. For example, from the figure, there are nodes with degree as high as 
$0.09$ and as low as $0.002$. The 
structure of the logistic attractor is actually characterized by ${{k} \over {N}}$ 
values $> \epsilon_c$.  

\section{\label{sec:level1}APPLICATIONS}
So far, we have been discussing the construction and analysis of RN from chaotic 
time series. It is also important to know how effectively our scheme can be applied to 
time series data from the real world. An important difference is that for standard 
chaotic systems, the dimensionality of the system is known \emph {a priori} and 
$M$ can be fixed accordingly while for real world data, this information is absent. 
In the conventional nonlinear time series analysis, one computes dynamical 
invariants as a function of $M$ and check for saturation with respect to $M$. 
For RN analysis, what is normally done is to use a sufficiently large value of 
$M$ to ensure a proper embedding of the underlying attractor \cite {gao1}. 
Our scheme indicates that a large $M$ requires a correspondingly large $N$ and 
$\epsilon_c$. However, the 
number of data points in real world time series is normally less (say, $< 10000$) and 
contaminated by different types of noise. One of the advantages of RN analysis is 
that it is possible to get information regarding the underlying system from a time 
series with limited number of data points. Hence for practical implementation of 
the scheme for real world data, we suggest that it is better to 
start the computation taking a small $M$, go for higher dimensions successively  
and check for saturation of the network measures for two successive dimensions, 
rather than using a very high embedding dimension to start with. 

To illustrate the potential applications of our approach, we consider two examples.
In the first example, we show that the proposed scheme is capable of identifying the 
dimensionality of the underlying system and the presence of white noise in real 
world data. For this, we present the RN analysis of the light 
curves from a black hole system GRS 1915+105. The light curves from this black hole 
system have been classified into $12$ spectroscopic states by Belloni et al. 
\cite {bell} and we take light curves from two representative states $\theta$ and 
$\chi$, which are shown in Fig.~\ref{f.16}. 
In an earlier paper \cite {kph2}, we have shown by computing $D_2$ that the state $\theta$ 
has signatures of deterministic nonlinear behavior (with $D_2 < 3$) and $\chi$ is white noise.  
We construct the RN from the two time series for different values of $M$ starting from 
$M = 2$ using the $\epsilon_c$ corresponding to each $M$ and compute the network measures 
in each case. We find that if the system is of finite dimension, 
the measures saturate beyond a certain $M$ which is taken as the 
dimension of the system. In Fig.~\ref{f.17}, 
we show the rescaled degree distributions of the two light curves for $M = 2, 3$ and $4$.  
Note that the degree distributions for the two states 
are completely different. For the $\chi$ state, they are Poissonian and shows the typical 
shift without any saturation  as $M$ increases, indicating pure white 
noise. On the other hand, the  
state $\theta$ is qualitatively different with the degree distributions getting  
saturated for $M=3$ and $4$ and remains stationary for any higher embedding dimension. 

Due to the inherent non-subjectivity in the choice of $\epsilon_c$, the scheme is also ideal 
for the surrogate analysis using network measures CC and CPL as discriminating statistic to 
detect deterministic nonlinearity in real world data. It can be used as complementary to 
conventional analysis with measures like $D_2$ and $K_2$ and has the added advantage that 
the length of the data required is much less compared to conventional methods.  

\begin{figure}
\includegraphics[width=0.96\columnwidth]{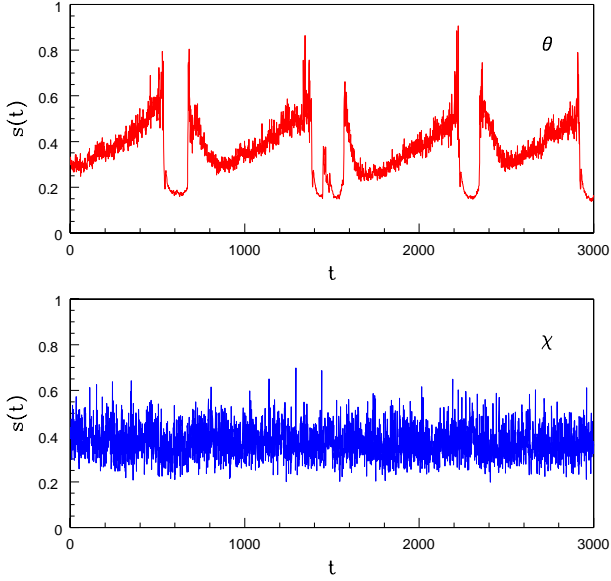}%
\caption{\label{f.16}(Color online) Part of the astrophysical light curves from the black 
hole system GRS 1915 + 105 for two temporal states, $\theta$ and $\chi$ 
(Ref: xte.mit.edu/~ehm/1915\_lightcurves.html).} 
\label{f.16}
\end{figure}

\begin{figure}
\includegraphics[width=0.96\columnwidth]{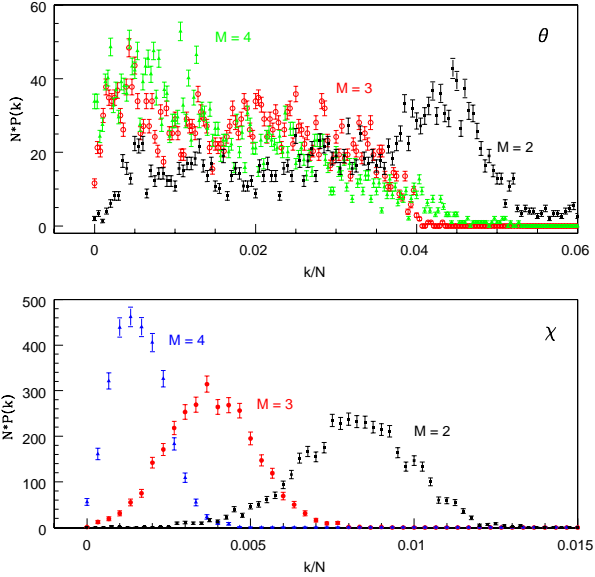}%
\caption{\label{f.17}(Color online) Rescaled degree distributions computed from the RNs for 
the two light curves $\theta$ and $\chi$ for three $M$ values. The distributions are for 
$M = 2$ (filled black squares), $M = 3$ (open red circles) 
and $M = 4$ (filled green triangles appearing in light gray shade in print) for $\theta$ state. 
The distributions for the three $M$ values for the $\chi$ state can be clearly distinguished, 
as indicated. In both cases, $N = 3000$. Note that for the $\theta$ 
state the two degree distributions for $M=3$ and $4$ almost coincide.} 
\label{f.17}
\end{figure}

As the second application, we consider detection of a dynamical transition using the 
RN measures derived from our method. This is known to be an important application of 
RNs \cite {dong1,kurt}.  The example we choose  
is the chaos-hyperchaos transition in a time delayed system, namely, the 
Mackey-Glass (M-G) system \cite {mack} given by the equation:
\begin{equation}
    {{dx} \over {dt}} = {{\beta x_{\tau_{D}}} \over {1+x_{\tau_D}^n}} - \gamma x 
    \label{eq:18}
\end{equation} 
The constants $\beta, \gamma, \tau_D$ and $n$ are real numbers and $x_{\tau_D}$ represents the value of the 
variable $x$ at time $t - \tau_D$. Depending on the values of the parameters, the system displays a range of 
periodic, chaotic and hyperchaotic dynamics. We fix the parameters $\beta = 2, \gamma = 1, n= 10$ with 
$\tau_D$ as the control parameter. As the value of the parameter $\tau_D$ increases, the asymptotic state of 
the system changes from periodic to chaotic and then to hyperchaos. We have recently shown \cite {kph5} 
using a dimensional analysis that the transition to hyperchaos occurs at a critical value 
$\tau_D = 4.038$. Here we check whether the RN measures obtained from our scheme can detect this 
transition.

\begin{figure}
\includegraphics[width=0.96\columnwidth]{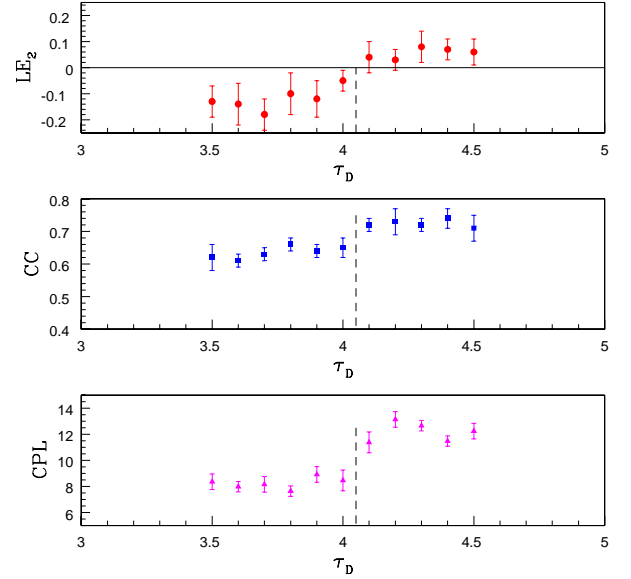}%
\caption{\label{f.18}(Color online) Variation of the CPL and CC of the RN constructed by our scheme 
from the time series of M-G system for the range of $\tau_D$ values representing the transition from 
chaos to hyperchaos. Corresponding values of the second largest LE (denoted $LE_2$) are 
shown in the top panel. The dashed vertical line indicates the transition point.} 
\label{f.18}
\end{figure}

To do the analysis, we first generate time series of length $100000$ from the system for $\tau_D$ values 
ranging from $3.5$ to $4.5$ increasing in steps of $0.1$. Each of this time series is then split up 
into $10$ time series of length $N_T = 10000$. We construct RN from all these and compute the measures 
CC and CPL taking $M = 4$ and $\epsilon = 0.14$, which is the minimum $M$ value required for a 
hyperchaotic attractor (any higher embedding dimension is equally good).   
Using the TISEAN package \cite {hegg}, we also 
compute the second largest Lyapunov Exponent (LE) (denoted $LE_2$) which crosses zero as the system 
cross over to hyperchaotic phase. The results of computations are shown in Fig.~\ref{f.18}, where 
the vertical dashed line indicates the transition point. The error bar comes from the standard 
deviation of the values obtained from the ten time series used for each $\tau_D$. It is clear that the 
present scheme can detect the transition as both CC and CPL show a discontinuity at the 
transition point.   

\section{\label{sec:level1}DISCUSSION AND CONCLUSION}
Recurrence networks have become important statistical tools for characterising the structural properties of chaotic attractors. They are complex networks constructed from chaotic time series using a suitable scheme that maps information inherent in the time series into the network domain. One can then use the network based measures to quantify the geometric properties of the underlying attractor. The distinct advantages of such an exercise are that the network measures can be efficiently computed from less number of data points in the time series and these measures can also be used as complimentary to conventional measures of nonlinear time series analysis. Here we present a general scheme to construct the RN from a chaotic time series that can be applied equally well to time series from standard chaotic attractors as well as real world data. We use identical value of threshold to construct the RN from different time series for a given embedding dimension. The scheme is thus found to be suitable to compare the structural properties of two chaotic attractors by computing the statistical measures of the corresponding recurrence networks. To illustrate how the scheme can be implemented in practice, it is used for the analysis of light curves from a black hole system and to identify the transitions between two dynamical regimes in a time delayed system. 

It is important to know the merits, limitations and potential applications of the proposed scheme 
for its proper implementation. Several methods have already been proposed in the literature for 
the construction of RN, 
as discussed in \S I. These methods mainly differ in the criteria for the selection of $\epsilon_c$. 
The main aspects in our approach compared to the earlier methods are the uniform deviate transformation  
of the time series and the criterion used for the selection of $\epsilon_c$ with its linkage to embedding 
dimension $M$. These changes make it possible to look for a uniform critical threshold for different 
chaotic time series.

We do not claim that the RN constructed by our approach is optimum for all the different types of 
time series and applications. The critical range $\Delta \epsilon$ presented in Table 1 for each 
$M$ is an empirical choice resulting from the analysis on a limited number of nodes $(N < 10000)$. 
It is primarily motivated to get a uniform range for different chaotic attractors as explained 
in \S III and is not a rigorous result. Though we expect the range to be valid in general, it may 
require improvement for specific time series and more accurate applications. However, by getting an 
identical value of $\epsilon_c$ for different time series, we are able to achieve a certain level of 
non subjectiveness in the construction of RN, especially for the analysis of time series from the 
real world, as we have shown explicitly. 

Another important outcome of the present approach not reported previously, is the realisation 
that the value of $\epsilon_c$ should be linked to $M$. This implies that the choice of $M$ is 
equally important for RN construction from time series and a very large value of $M$, 
as is generally believed, may not 
provide optimum result with limited number of data points. It is also interesting to note that 
there is always a part in the degree distribution of the RN from a chaotic attractor that 
corresponds to Poisson distribution where the $k$ values occur more by chance than by choice. 
Its position in the degree distribution depends on the choice of $\epsilon_c$ and $M$.  

There are at least four important potential applications for our approach, three of which we 
have shown here explicitly: 

i) to compare the structural properties of two chaotic attractors using network measures through 
the construction of RN 

ii) to study the transition between two dynamical regimes as a control parameter is varied 

iii) to identify the dimension of the underlying attractor from the analysis of time series data and 

iv) for surrogate analysis using any of the RN measures as discriminating statistic where a 
non subjective comparison of the network measures from data and surrogates is required. 
 
Finally, an important step forward in our analysis is to try and develop a similar scheme for 
RNs where the connections have weight factors. Here we have considered unweighted 
RNs so that the resulting adjacency matrix is binary. We hope that a weighted RN 
can unravel more information regarding the topological and structural properties 
of chaotic attractors. Another possible application of the scheme, that is 
important in the analysis of real world data, is to study the effect of noise on 
RN and the measures derived from it. These works are currently in progress and will be 
reported elsewhere.

\begin{acknowledgments}
RJ and KPH acknowledge the financial support from the Science and Engineering Research Board (SERB),  
Govt. of India, in the form of a Research Project No. SR/S2/HEP-27/2012. 
RJ and KPH acknowledge the computing facilities in IUCAA, Pune.
\end{acknowledgments}

\end{document}